\documentclass[aip,graphicx,groupedaddress]{revtex4-1}

\usepackage{graphicx}
\usepackage{amsmath}
\usepackage{amsbsy}
\usepackage{dcolumn}
\usepackage{bm}
\usepackage{braket}
\draft 

\begin{document}


\title{Spin-wave-induced spin torque in Rashba spin-orbit coupling system} 



\author{Nobuyuki Umetsu}
\affiliation{Department of Applied Physics, Tohoku University, Aoba 6-6-05, Aoba-ku, Sendai 980-8579, Japan}
\author{Daisuke Miura}
\affiliation{Department of Applied Physics, Tohoku University, Aoba 6-6-05, Aoba-ku, Sendai 980-8579, Japan}
\author{Akimasa Sakuma}
\affiliation{Department of Applied Physics, Tohoku University, Aoba 6-6-05, Aoba-ku, Sendai 980-8579, Japan}


\date{\today}

\begin{abstract}
We study the effects of Rashba spin-orbit coupling on the spin torque induced by spin waves, which are the plane wave dynamics of magnetization.
The spin torque is derived from linear response theory,
and we calculate the dynamic spin torque by considering the impurity-ladder-sum vertex corrections.
This dynamic spin torque is divided into three terms: a damping term, a $distortion$ term, and a correction term for the equation of motion.
The $distorting$ torque describes a phenomenon unique to the Rashba spin-orbit coupling system,
where the distorted motion of magnetization precession is subjected to the anisotropic force from the Rashba coupling.
The oscillation mode of the precession exhibits an elliptical trajectory,
and the ellipticity depends on the strength of the nesting effects,
which could be reduced by decreasing the electron lifetime.

\end{abstract}


\maketitle 

\section{Introduction}

The space-dependent spin dynamics in ferromagnetic metals have been the subject of considerable investigation over the past several years.
This subject is of great significance for technological applications such as spin-transfer torque random-access memory (STT-RAM) and microwave generators or detectors.
In the spintronics devices, which are nanoscale magnetic multilayer systems,
inhomogeneous magnetization dynamics are caused by an external applied field or thermal excitations
due to the finite size effects\cite{Shaw2009,Nembach2013} and the interface effects \cite{UrbanWoltersdorfHeinrich2001}.
The spin torques, which arise in non-equilibrium conditions include some terms with different forms \cite{ZhangLi2004,Kohno2006,KohnoShibat2007,TataraKohnoShibataLemahoLee2007,SakaiKohno2014},
and
the space-dependent spin torques caused by inhomogeneous magnetization dynamics are different from the current-induced spin torques,
namely, the spin transfer torque\cite{Slonczewski1996,BazaliyJonesZhang1998,BarnesMaekawa2005,XiaoZangwillStiles2006,TserkovnyakSkadsemBrataasBauer2006} and the spin-orbit torque
\cite{ObataTatara2008,PiWonBaeLeeChoKimSeo2010,MihaiGaudinAuffretRodmacqSchuhlPizziniVogelGambardella2010,LiuPaiLiTsengRalphBuhrman2012,KhvalkovskiyCrosApalkovNikitinKrounbiZvezdinAnaneGrollierFert2013,LinderAlidoust2013}.
The former torque is due to the indirect interaction between magnetization through conduction electrons,
while the latter torque is due to the direct interaction between magnetization and conduction electrons.
Generally, the space-dependent torque acting on magnetization at $\bm{r}$ can be represented by
$\int{\rm d}\bm{r}'{\rm d}t\,\bm{M}(\bm{r},t)\times\tilde{\chi}(\bm{r},\bm{r}',t,t')\bm{M}(\bm{r}',t')$,
where $\tilde{\chi}(\bm{r},\bm{r}',t,t')$ is the spin susceptibility tensor (in detail, see \eqref{eq:chilm}).
We are interested in the dynamic part of this torque,
which describes the interaction between magnetization $\bm{M}\left(\bm{r}\right)$
and the time-derivative of other magnetization $\dot{\bm{M}}\left(\bm{r}'\right)$,
and it corresponds to Gilbert damping torque $\alpha\bm{M}\times\dot{\bm{M}}$ when magnetization precesses uniformly,
where $\alpha$ is the Gilbert damping constant\cite{Gilbert2004}.
Although a large number of studies have considered magnetization damping in uniform precession systems
\cite{Siimanek2003,TserkovnyakBrataasBauerHalperinI.2005,Kohno2006,Duine2007,Kambersky2007,Gilmore2007,Garate2009} (i.e., local damping),
little is known about the damping mechanism in inhomogeneous magnetization dynamics\cite{KorenmanPrange1972,Tserkovnyak2009,UmetsuMiuraSakuma2012a,Sakuma2015} (i.e., nonlocal damping).
Owing to the development of recent experimental techniques, detailed and unified theories of nonlocal damping are more desirable for spintronics applications.

Previous theoretical studies show that magnetic damping arising from spin diffusion
is enhanced by the spin wave
that describes the plane wave dynamics of magnetization\cite{Tserkovnyak2009}.
The damping coefficient is represented by $\alpha=\alpha_0+\eta q^2$, especially in the long wavelength limit,
where the first term represents damping for uniform precession systems and
the second term is the contribution from spin wave motion with wave vector $\bm{q}$.
The coefficient $\eta$ is the diffusion constant, which depends strongly on the lifetime of the conduction electrons.
Moreover,  experimental evidence for nonlocal damping has been reported by the recent work \cite{Nembach2013}, where
the Gilbert damping constants for two modes (center mode and edge mode)
depend on the size of the nanoscale magnets, implying that additional damping torque was caused by finite size effects.

The purpose of our work is to calculate the dynamic part of the spin torque
induced by the spin wave motion of magnetization in a Rashba spin-orbit coupling (RSOC) system.
No previous studies have discussed the inhomogeneous magnetization dynamics in SOC systems.
In our previous work\cite{UmetsuMiuraSakuma2015},
we discuss qualitatively the effects of SOC in the RSOC system without electrons scattering by impurities. 
In this paper, we show the results including effects of the impurity scattering beyond the previous results in the clean limit
and discuss the effects of  impurity-ladder-sum vertex corrections.

The reminder of this paper is organized as follows.
In Sec. II, we explain our model and formulation.
In Sec. III, we show our calculation results for the dynamic part of the spin-wave-induced spin torque.
Finally, a summary of our work is given in Sec. IV.

\section{Model and Formulation}

The exchange coupling between electrons and magnetization is represented by
\begin{equation}
\mathcal{H}_{{\rm ex}}\left( t\right)=-2\Delta\int{\rm d}\bm{r}\bm{n}\left(\bm{r},t\right)\cdot\bm{s}\left(\bm{r}\right),
\end{equation}
where $\bm{n}\left(\bm{r},t\right)=\bm{M}\left(\bm{r},t\right)/M$ is the unit vector of the time-dependent magnetization at $\bm{r}$ and
$\Delta$ is the exchange energy. Here, $\bm{s}\left(\bm{r}\right)=\psi^{\dagger}\left(\bm{r}\right)\hat{\bm{\sigma}}\psi\left(\bm{r}\right)/2$
is the electron spin density with the electron field operator
$\psi^{\left(\dagger\right)}\left(\bm{r}\right)=[\psi^{(\dagger)}_\uparrow\left(\bm{r}\right),\psi^{(\dagger)}_\downarrow\left(\bm{r}\right)]$
and Pauli matrix vector $\hat{\bm{\sigma}}$  (the hat symbol indicates a $2\times 2$ matrix).
The direction of magnetization is parallel to $z$-axis at the equilibrium state,
and the contributions from the transverse component (denoted by $\perp$) are treated perturbatively.
Based on linear response theory, the time-dependent energy is written as
\begin{align}
&E_{\rm ex}\left(t \right)=\frac{1}{{\rm i}\hbar}\int_{-\infty}^{t}{\rm d}t'\left\langle \left[\mathcal{H}_{{\rm ex}}\left(t\right),\mathcal{H}_{{\rm ex}}\left(t'\right)\right]\right\rangle \nonumber \\
=&-4\Delta^2N\int{\rm d}t'\int{\rm d}\bm{r}' \bm{n}^\perp\left(\bm{r},t\right)\tilde{\chi}\left(\bm{r},t,\bm{r}',t'\right) \bm{n}^\perp\left(\bm{r}',t'\right),
\end{align}
where $N$ is the number of unit cells. The spin susceptibility tensor $\tilde{\chi}$ contains elements  $(l,m)$, which are given by
\begin{equation}
\chi^{lm}\left(\bm{r},t,\bm{r}',t'\right)=-N^{-1}\frac{1}{{\rm i}\hbar}\left\langle \left[s^{l}\left(\bm{r},t\right),s^{m}\left(\bm{r}',t'\right)\right]\right\rangle \Theta\left(t-t'\right).
\label{eq:chilm}
\end{equation}
The effective field is derived from
$\bm{H}_{\rm ex}\left( t \right)=M^{-1}E_{\rm ex}\left(t \right)/\partial \bm{n}\left(\bm{r},t\right)$,
and the spin torque is obtained by
$\bm{T}_{\rm st}\left(\bm{r},t\right)=-\gamma \bm{n}\left(\bm{r},t\right)\times\bm{H}_{\rm ex}\left(t\right)$,
where $\gamma$ is the gyromagnetic ratio.
In the first approximation, neglecting the products of perturbed quantities,
the spin torque is written as
\begin{equation}
\bm{T}_{\rm st}\left(\bm{r},t\right)=\gamma\frac{2\Delta}{M}\bm{n}^z\times\left<\bm{s}^\perp\left(\bm{r},t\right)\right>,
\end{equation}
where $\left\langle \bm{s}^{\perp}\left(\bm{r},t\right)\right\rangle
=2\Delta \int{\rm d}t' \int{\rm d}\bm{r}' \tilde{\chi}\left(\bm{r},t,\bm{r}',t'\right) \bm{n}^\perp\left(\bm{r}',t'\right)$
is the transverse component of the electron spin density expectation value.
In slow modulations of the magnetization motion,
$\bm{n}^{\perp}\left(\bm{r}',t'\right)\approx\bm{n}^{\perp}\left(\bm{r}',t\right)+\left(t'-t\right)\dot{\bm{n}}^{\perp}\left(\bm{r}',t\right)$
is satisfied, and the spin torque is rewritten as
\begin{equation}
\bm{T}_{\rm st}\left(\bm{r},t\right)\approx
\int{\rm d}\bm{r}'\bm{n}^z\times
\left[\tilde{\beta}\left(\bm{r}-\bm{r}'\right)\bm{n}^{\perp}\left(\bm{r}',t\right)
+\tilde{\alpha}\left(\bm{r}-\bm{r}'\right)\dot{\bm{n}}^{\perp}\left(\bm{r}',t\right)
\right],
\label{eq:Tst}
\end{equation}
in an electron system with translation symmetry.
Here,
\begin{equation}
\tilde{\beta}\left(\bm{r}\right)=\frac{4\gamma\Delta^{2}N}{M}
\tilde{\chi}\left(\bm{r},\omega=0\right),
\end{equation}
is the coefficient tensor of the static part, and
\begin{equation}
\tilde{\alpha}\left(\bm{r}\right)
={\rm i}\frac{4\gamma\Delta^{2}N}{M}
\lim_{\omega \to 0}\frac{\partial}{\partial\omega}\tilde{\chi}\left(\bm{r},\omega\right),
\end{equation}
is the coefficient tensor of the dynamic part.
The Fourier transformation of spin susceptibility is defined as
$\tilde{\chi}\left(\omega\right)=\int{{\rm d}t}{\rm e}^{{\rm i}\omega t}\tilde{\chi}\left(t\right)$.
The first term on the right-hand side of Eq.\eqref{eq:Tst} is the RKKY-type interaction,
which is regarded as the time-dependent correction for precession torque caused by the static magnetic field; however,
this term is irrelevant for our work.
In this study, we focus only on the dynamic  spin torque, which is described in second term on the right-hand side of Eq.\eqref{eq:Tst}.
Expanding the magnetization motion using a plane wave with wave vector $\bm{q}$,
the dynamic part of $\bm{T}_{\rm st}$ in $\bm{q}$-space is written as $\bm{n}^z\times\tilde{\alpha}_{\bm{q}}\dot{\bm{n}}^{\perp}_{\bm{q}}$.
Here,
\begin{equation}
\tilde{\alpha}_{\bm{q}}=\frac{4\Delta^2}{\hbar S}
\lim_{\omega \to 0}\frac{\partial}{\partial \omega}{\rm Im}\tilde{\chi}_{\bm{q}}\left(\omega\right),
\label{tildealpha}
\end{equation}
where $\hbar S$ is the spin angular momentum of magnetization in a unit volume $v=V/N$.
Using a simple calculation, it can be shown that the $\omega$-derivative of the real part of $\tilde{\chi}_{\bm{q}}=\int{\rm d}\bm{r}\tilde{\chi}\left(\bm{r}\right){\rm e}^{-{\rm i}\bm{q}\cdot\bm{r}}$
vanishes in $\omega\to 0$.
Hereafter, the matrix elements related to the $z$-component of $\tilde{\alpha}_{\bm{q}}$ are neglected
because we are not concerned with the longitudinal motion in this study.

We calculate $\tilde{\chi}_{\bm{q}}\left(\omega\right)$ using a Green function technique
that considers the electron-impurity (spin-independent) interaction within the first Born approximation \cite{Garate2009}.
The spin susceptibility is obtained using analytic continuation of the Matsubara Green's function, which is written as
\begin{align}
&\chi_{\bm{q}}^{lm}\left({\rm i}\nu_{n}\right)
=\int_{0}^{1/T}{\rm d}{\eta}{\rm e}^{{\rm i}\nu_{n}\eta}\left\langle {\rm T}s_{\bm{q}}^{l}\left(\eta\right)s_{-\bm{q}}^{m}\right\rangle \nonumber \\
=&-\frac{T}{4}\sum_{\bm{k}}\sum_{n'}{\rm Tr}
\Bigl[\hat{g}\left({\rm i}\omega_{n'},\bm{k}\right)
\hat{\Gamma}^l\left({\rm i}\omega_{n'},{\rm i}\omega_{n'}+{\rm i}\nu_n,\bm{q}\right)
\hat{g}\left({\rm i}\omega_{n'}+{\rm i}\nu_{n},\bm{k}+\bm{q}\right)\hat{\sigma}^m
\Bigr],
\label{eq:chilm}
\end{align}
where $\hat{g}\left({\rm i}\omega_{n},\bm{k}\right)=-\int^{1/T}_0{\rm d}{\eta}{\rm e}^{{\rm i}\omega_{n}\eta} \left<{\rm T}\psi_{\bm{k}}\left(\eta\right)\psi^\dagger_{\bm{k}}\right>$
is the one-particle Green's function and
$\hat{\Gamma}^l\left({\rm i}\omega_{n'},{\rm i}\omega_{n'}+{\rm i}\nu_n,\bm{q}\right)$
is the vertex function.
This vertex function satisfies the following Ward identity:
\begin{align}
\Gamma_{\sigma_1\sigma_2}^{l}
&\left({\rm i}\omega_{n'},{\rm i}\omega_{n'}+{\rm i}\nu_n,\bm{q}\right)
=\, \sigma_{\sigma_1\sigma_2}^{\mu} \nonumber \\
&+\frac{\hbar/\tau}{\epsilon_{\rm F}}\sum_{\rho_{1}\rho_{2}}\Pi_{\sigma_1\rho_{1}\rho_{2}\sigma_2}
\left({\rm i}\omega_{n'},{\rm i}\omega_{n'}+{\rm i}\nu_n,\bm{q}\right)
\Gamma_{\rho_{1}\rho_{2}}^{l}\left({\rm i}\omega_{n'},{\rm i}\omega_{n'}+{\rm i}\nu_n,\bm{q}\right),
\label{eq:vertex}
\end{align}
where
\begin{equation}
\Pi_{\sigma_1\rho_{1}\rho_{2}\sigma_2}\left({\rm i}\omega_{n'},{\rm i}\omega_{n'}+{\rm i}\nu_n,\bm{q}\right)
=\frac{\epsilon_{{\rm F}}}{\pi\nu_{{\rm F}}N}\sum_{\bm{k}}g_{\sigma_1\rho_{1}}\left({\rm i}\omega_{n'}+{\rm i}\nu_n,\bm{k}+\bm{q}\right)
g_{\rho_{2}\sigma_2}\left({\rm i}\omega_{n'},\bm{k}\right).
\label{eq:Pi_element}
\end{equation}
Here, $\hbar/\tau=\pi\nu_{\rm F} n_0u^2$ is the inverse electron lifetime $\tau$
[$\nu_{\rm F}$ is the density of states per volume at the Fermi level, $n_0$ is the density of impurities, and
$u$ is the scattering constant of the short range potential $uv\delta(\bm{r})$].
We define the following $4\times 4$ matrices:
\begin{equation}
\check{K}=\left(1-\frac{\hbar/\tau}{\epsilon_{\rm F}}\check{\Pi}\right)^{-1},\qquad
\check{\Pi}=\left(\begin{array}{cccc}
\Pi_{\uparrow\uparrow\uparrow\uparrow} & \Pi_{\uparrow\uparrow\downarrow\uparrow} & \Pi_{\uparrow\downarrow\uparrow\uparrow} & \Pi_{\uparrow\downarrow\downarrow\uparrow}\\
\Pi_{\uparrow\uparrow\uparrow\downarrow} & \Pi_{\uparrow\uparrow\downarrow\downarrow} & \Pi_{\uparrow\downarrow\uparrow\downarrow} & \Pi_{\uparrow\downarrow\downarrow\downarrow}\\
\Pi_{\downarrow\uparrow\uparrow\uparrow} & \Pi_{\downarrow\uparrow\downarrow\uparrow} & \Pi_{\downarrow\downarrow\uparrow\uparrow} & \Pi_{\downarrow\downarrow\downarrow\uparrow}\\
\Pi_{\downarrow\uparrow\uparrow\downarrow} & \Pi_{\downarrow\uparrow\downarrow\downarrow} & \Pi_{\downarrow\downarrow\uparrow\downarrow} & \Pi_{\downarrow\downarrow\downarrow\downarrow}
\end{array}\right).
\label{eq:K_matrix}
\end{equation}
Then, the vertex function is given by
\begin{align}
\Gamma_{\sigma_1\sigma_2}^{l}\left({\rm i}\omega_{n'},{\rm i}\omega_{n'}+{\rm i}\nu_n,\bm{q}\right) & =\sum_{\rho_{1}\rho_{2}}K_{\sigma_1\rho_{1}\rho_{2}\sigma_2}\left({\rm i}\omega_{n'},{\rm i}\omega_{n'}+{\rm i}\nu_n,\bm{q}\right)
\sigma_{\rho_{1}\rho_{2}}^{l}.
\label{eq:vertex2}
\end{align}
Substituting Eq.\eqref{eq:vertex2} into Eq.\eqref{eq:chilm} and
performing the conventional $\bm{k}$-sum and $n'$-sum in the low-temperature limit,
we obtain the following form of Eq.\eqref{tildealpha}:
\begin{align}
\tilde{\alpha}_{\bm{q}}
&=\alpha_{\bm{q}}\hat{\sigma}^{0}+\alpha^{x}_{\bm{q}}\hat{\sigma}^{x}
+\alpha^{y}_{\bm{q}}(-{\rm i})\hat{\sigma}^y+\alpha^{z}_{\bm{q}}\hat{\sigma}^z,
\nonumber \\
\alpha_{\bm{q}}&=\frac{\nu_{{\rm F}}\Delta^{2}}{2S\epsilon_{{\rm F}}}{\rm Re}\sum_\sigma
\left(\check{K}_{\bm{q}}^{{\rm AR}}\check{\Pi}_{\bm{q}}^{{\rm AR}}-\check{K}_{\bm{q}}^{{\rm AA}}\check{\Pi}_{\bm{q}}^{{\rm AA}}\right)_{\sigma\sigma\bar{\sigma}\bar{\sigma}},
\nonumber \\
\alpha_{\bm{q}}^x&=\frac{\nu_{{\rm F}}\Delta^{2}}{2S\epsilon_{{\rm F}}}{\rm Im}\sum_\sigma \sigma
\left(\check{K}_{\bm{q}}^{{\rm AR}}\check{\Pi}_{\bm{q}}^{{\rm AR}}-\check{K}_{\bm{q}}^{{\rm AA}}\check{\Pi}_{\bm{q}}^{{\rm AA}}\right)_{\bar{\sigma}\sigma\bar{\sigma}\sigma},
\nonumber \\
\alpha_{\bm{q}}^y&=\frac{\nu_{{\rm F}}\Delta^{2}}{2S\epsilon_{{\rm F}}}{\rm Im}\sum_\sigma \sigma
\left(\check{K}_{\bm{q}}^{{\rm AR}}\check{\Pi}_{\bm{q}}^{{\rm AR}}-\check{K}_{\bm{q}}^{{\rm AA}}\check{\Pi}_{\bm{q}}^{{\rm AA}}\right)_{\sigma\sigma\bar{\sigma}\bar{\sigma}},
\nonumber \\
\alpha_{\bm{q}}^z&=\frac{\nu_{{\rm F}}\Delta^{2}}{2S\epsilon_{{\rm F}}}{\rm Re}\sum_\sigma
\left(\check{K}_{\bm{q}}^{{\rm AR}}\check{\Pi}_{\bm{q}}^{{\rm AR}}-\check{K}_{\bm{q}}^{{\rm AA}}\check{\Pi}_{\bm{q}}^{{\rm AA}}\right)_{\bar{\sigma}\sigma\bar{\sigma}\sigma},
\label{eq:alpha_tilde}
\end{align}
where $(\check{\Pi}_{\bm{q}}^{\rm AR, AA})_{\sigma_1\rho_1\rho_2\sigma_2}
=(\epsilon_{\rm F}/\pi\nu_{\rm F}N)\sum_{\bm{k}}g^{\rm A}_{\sigma_1\rho_1}(\omega=0,\bm{k}+\bm{q})g^{\rm R,A}_{\rho_2\sigma_2}(\omega=0,\bm{k})$.
The superscripts $\rm R$ and $\rm A$ denote the retarded Green's function and the advanced Green's function, respectively.

In Eq.\eqref{eq:alpha_tilde}, $\alpha_{\bm{q}}$ is the conventional Gilbert damping coefficient, and
$\alpha_{\bm{q}}^y$, which is the coefficient of $\bm{n}^z\times(-{\rm i})\hat{\sigma}^y\dot{\bm{n}}^\perp_{\bm{q}}=-\dot{\bm{n}}^\perp_{\bm{q}}$,
is the correction for the equation of motion.
Neither $\alpha_{\bm{q}}^x$ nor $\alpha_{\bm{q}}^y$ have been examined in previous studies
because these terms disappear in non-SOC systems and in uniform precession systems.
The physical meaning of these terms is discussed in the next section.

The electron system of our model is a 2D electron gas incorporating RSOC\cite{Garate2009},
which is described by
\begin{equation}
\mathcal{H}=\epsilon_{{\rm F}}\sum_{\bm{k}}\psi_{\bm{k}}^{\dagger}\left(\left|\bm{k}_{0}\right|^{2}+\bm{\Lambda}_{\bm{k}}\cdot\hat{\bm{\sigma}}\right)\psi_{\bm{k}},
\end{equation}
where $\epsilon_{{\rm F}}=\hbar^2 k_{{\rm F}}^{2}/2m$ is the Fermi
energy and $\bm{k}_{0}=\left(k_{x}/k_{{\rm F}},k_{y}/k_{{\rm F}}\right)$
is the normalized wave vector.
Here, $\bm{\Lambda}_{\bm{k}}$ is defined as
\begin{equation}
\bm{\Lambda}_{\bm{k}}=\lambda_{\bm{k}_{0}}\left(\sin\theta_{\bm{k}}\cos\phi_{\bm{k}},\sin\theta_{\bm{k}}\sin\phi_{\bm{k}},\cos\theta_{\bm{k}}\right),
\end{equation}
where $\phi_{\bm{k}}=-\tan^{-1}\left(k_{x}/k_{y}\right)$, $\theta_{\bm{k}}=\cos^{-1}\left(-\Delta_{0}/\lambda_{\bm{k}_{0}}\right)$, and $\lambda_{\bm{k}_{0}}=\sqrt{\lambda^{2}k_{0}^{2}+\Delta_{0}^{2}}$.
Moreover, $\Delta_{0}=\Delta/\epsilon_{{\rm F}}$, and $\lambda$ is the strength of RSOC
(the conventional Rashba parameter $\alpha_{\rm R}$ is written as $\alpha_{\rm R}=\lambda\epsilon_{\rm F}/k_{\rm F}$).
The eigenstate corresponding to the eigenenergy of $\mathcal{H}$,
$\epsilon_{\bm{k}\pm}=\epsilon_{{\rm F}}\left(k_{0}^{2}\pm\lambda_{\bm{k}_{0}}\right)$, is given by
\begin{equation}
\ket{{\bm{k}},+(-)}  ={\rm e}^{-(+){\rm i}\frac{\phi_{\bm{k}}}{2}}\cos\frac{\theta_{\bm{k}}}{2}\ket{\uparrow(\downarrow)}
+(-){\rm e}^{+(-){\rm i}\frac{\phi_{\bm{k}}}{2}}{\rm sin}\frac{\theta_{\bm{k}}}{2}\ket{\downarrow(\uparrow)}.
\label{eq:eigenstate}
\end{equation}
Then, the matrix elements of the retarded and advanced Green's function are written as
\begin{align}
g_{\sigma_{1}\sigma_{2}}^{\rm R,A }\left({\rm i}\omega_n,\bm{k}\right)
&=\sum_{\alpha=\pm}\braket{\sigma_1|\alpha}_{\bm{k}}g^{\rm R,A}_{\alpha}\left({\rm i}\omega_n,\bm{k}\right)\braket{\alpha|\sigma_{2}}_{\bm{k}},
\nonumber \\
g^{\rm R, A}_{\alpha}\left({\rm i}\omega_{n},\bm{k}\right)
&=\left({\rm i}\omega_{n}+\epsilon_{{\rm F}}-\epsilon_{\bm{k}\alpha}\pm{\rm i}\hbar/\tau \right)^{-1}
,\qquad\left({\rm +\,for\,R, -\, for\, A}\right).
\label{eq:matrix_g_element}
\end{align}
We can  calculate $\tilde{\alpha}_{\bm{q}}$ from Eq.\eqref{eq:K_matrix}, Eq.\eqref{eq:vertex2},
Eq.\eqref{eq:alpha_tilde}, Eq.\eqref{eq:eigenstate}, and Eq.\eqref{eq:matrix_g_element}.

\section{Results and Discussion}

In the numerical calculations, we set the following parameters:
$\Delta_0=0.1$, $\nu_{\rm F}=1/\epsilon_{\rm F}$, $\lambda=0.3$, and $S=1$.
We show the numerical results of two cases: $\hbar/\tau=0.01\epsilon_{\rm F}$ and $\hbar/\tau=0.05\epsilon_{\rm F}$.
However, we do not show the detail results of $\alpha_{\bm{q}}^y$ in this section
because we have determined that this term is sufficiently small and can be neglected.
In our results, the maximum absolute value of $\alpha^y_{\bm q}$ for $\hbar/\tau=0.01\epsilon_{\rm F}$ is about 0.07
and that for $\hbar/\tau=0.05\epsilon_{\rm F}$ is about 0.06,
which satisfies $\alpha_{\bm{q}}^y\ll 1$ (i.e., $\alpha_{\bm{q}}^y \dot{\bm{n}}^\perp\ll \dot{\bm{n}}^\perp$).

\subsection{Damping torque \label{damping}}

\begin{figure}[h]
\centering{}\includegraphics[scale=1.2]{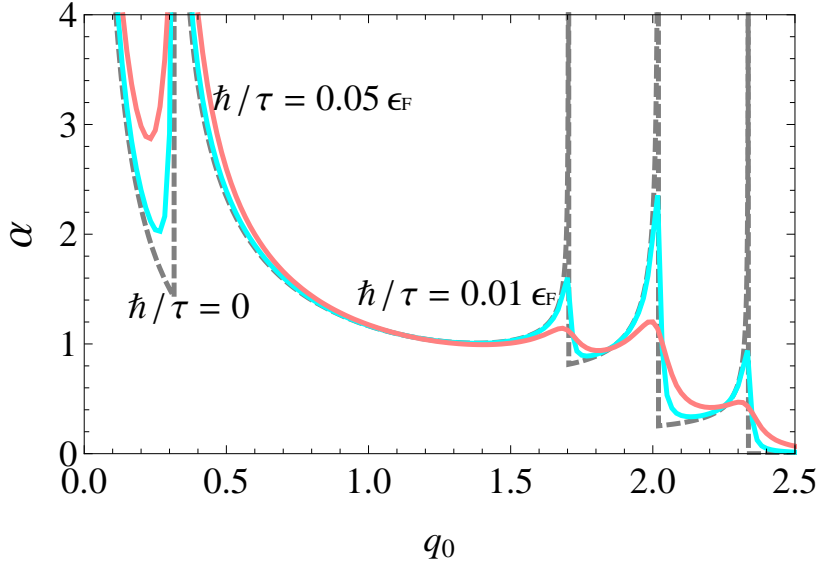}
\caption{The $q$-dependence of $\alpha_{q}$ (divided by $\nu_{\rm F}\Delta^2/S\epsilon_{\rm F}=10^{-2}$).
The results in the clean limit are indicated by the dashed line.
\label{fig:a}}
\end{figure}

In this subsection, we discuss the results of ${\alpha}_{\bm{q}}$, which is the coefficient of conventional damping torque, $\bm{n}^z\times\dot{\bm{n}}^\perp_{\bm{q}}$.
The $q$-dependence of $\alpha_{q}$ is shown in Fig.\ref{fig:a}.
It can be strictly shown that $\alpha_{\bm{q}}$ does not depend on the direction of $\bm{q}$.

In the clean limit,
\begin{equation}
\alpha_{q}
=\frac{\nu_{{\rm F}}\Delta^{2}}{S\epsilon_{{\rm F}}}\sum_{\alpha\beta}\frac{\left(k_{\alpha}^{2}-k_{\beta}^{2}\right)^{2}+\alpha\beta\lambda^{2}\left(k_{\alpha}^{2}+k_{\beta}^{2}\right)}{\left(k_{-}^{2}-k_{+}^{2}\right)^{2}}I_{\alpha\beta}\left(q_{0}\right),
\end{equation}
where
\begin{equation}
I_{\alpha\beta}\left(q_{0}\right)
=\frac{\Theta\left(q_{0}-\left|k_{\alpha}-k_{\beta}\right|\right)\Theta\left(k_{\alpha}+k_{\beta}-q_{0}\right)}{\sqrt{\left[\left(k_{\alpha}+k_{\beta}\right)^{2}-q_{0}^{2}\right]\left[q_{0}^{2}-\left(k_{\alpha}-k_{\beta}\right)^{2}\right]}},
\qquad \left(q_0=\frac{q}{k_{\rm F}}\right)
\label{eq:nesting}
\end{equation}
represents the strength of the nesting effects for spin excitations between the $\alpha$-band and 
the $\beta$-band. Here,
$k_{\pm}=\sqrt{1+\lambda^2/2\mp\sqrt{\lambda^2+\lambda^4/4+\Delta_0^2}}$
is the radius of the Fermi sphere of the $\pm$-band.
At $q_0=0,k_{-}\pm k_{+},2k_{\pm}$, $\alpha_q$ diverges because of the strong nesting caused by the adjoining Fermi surfaces.
However, in a non-SOC system, $\alpha_q$ diverges only at $q_0=k_-\pm k_+$ \cite{UmetsuMiuraSakuma2012a}
because intra-band transitions are forbidden (i.e., the contribution from $\alpha=\beta$ is zero).
The divergence at $q=0$ in SOC systems is well understood from previous results \cite{Kambersky2007},
but the divergence at $q_0=2k_{\pm}$ arising from the intra-band transitions is a new result
related to the inhomogeneous dynamics of magnetization.

From Fig.\ref{fig:a}, it is confirmed that the $\alpha_q$ divergence 
is suppressed by the impurity scattering of electrons, and
the value at each peak decreases with decreasing electron lifetimes.
These results imply that the nesting effects are reduced by broadening the Fermi level
due to the increase in self-energy.
There are cases that $\alpha_q$ increases with decreasing $\tau$
because the line width of each peak increases with $\tau$.
These behaviors are also confirmed in the non-SOC system in our previous study \cite{UmetsuMiuraSakuma2012a}.
This implies that $\alpha_q$ increases due to spin diffusion originating in inter-band transitions.

\begin{figure}[h]
\centering{}\includegraphics[scale=1.2]{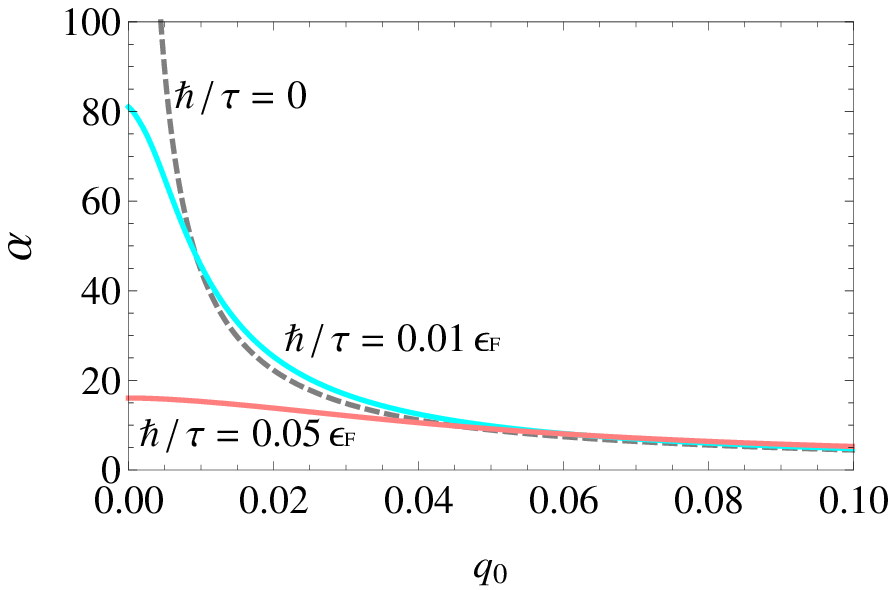}
\caption{Behavior of $\alpha_{q}$ near $q=0$.
\label{fig:a0}}
\end{figure}

The behaviors of $\alpha_{q}$ near $q=0$ are shown in Fig.\ref{fig:a0}.
The value of $\alpha_{q}$ at $q=0$ corresponding to the Gilbert damping constant in a uniform precession system
diverges in the clean limit, but it converges in the presence of impurities \cite{Kambersky2007}.
On the other hand, in a non-SOC system,
$\alpha_{q=0}$ is always zero regardless of the concentrations of impurities;
this behavior is required from the angular momenta conservation law, which is satisfied by taking the vertex correction (VC).
However, in the presence of $magnetic$ (spin-dependent) impurities,
$\alpha_{q=0}$ has a finite value due to the violation of the conservation law \cite{Kohno2006,UmetsuMiuraSakuma2012a}.
In a non-SOC system,
the spin-dependency of impurities causes dramatic differences in the $\alpha_q$ behavior;
however, the spin-dependence on impurities rarely appears in SOC systems\cite{Garate2009}.
Therefore, we do not need to provide a detailed analysis of the effects of magnetic impurities in our RSOC system.

\begin{figure}[h]
\centering{}\includegraphics[scale=1.2]{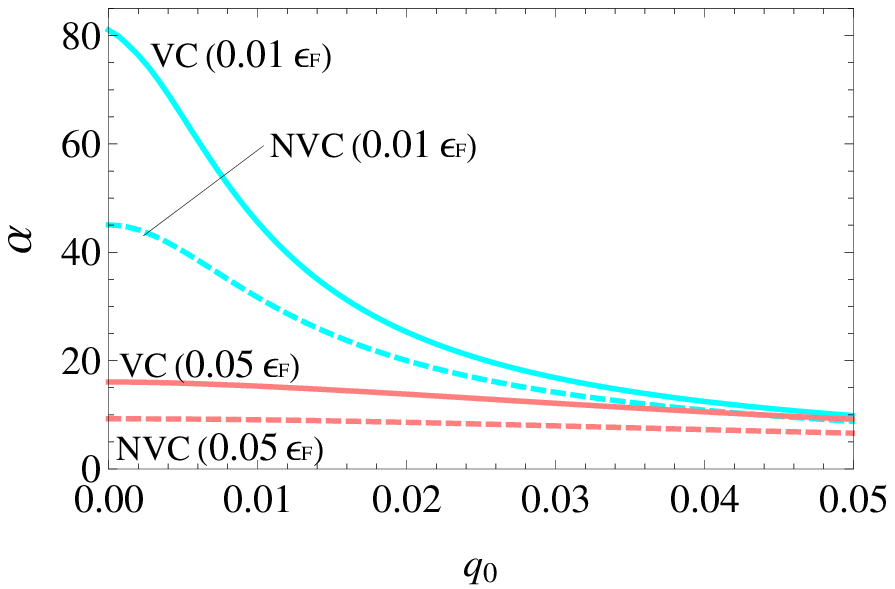}
\caption{Comparison of $\alpha_q$ in $q_0<0.05$ including the vertex corrections (VC) and not including the vertex corrections (NVC).
The solid lines show the results for VC, and the dashed lines show the results for NVC.
The blue lines are results for $\hbar/\tau=0.01\epsilon_{\rm F}$, and
the red lines are results for $\hbar/\tau=0.05\epsilon_{\rm F}$.
\label{fig:a_vc}}
\end{figure}

From Fig.\ref{fig:a_vc}, it is clear that results including the VC term are larger than those without the VC term, especially at $q\simeq 0$.
However, in a non-SOC system,
$\alpha_q$ with VC is zero at $q=0$,
and $\alpha_q$ is smaller than the results without VC.
The VC term reduces the overvalue of the contributions from spin diffusion originating in inter-band transitions.
In an SOC system at $q\simeq 0$, the contributions from intra-band transitions are much larger than those from inter-band transitions.
We conclude that the VC term of an SOC system enhances the contributions from nesting effects, which originate in intra-band transitions.

\subsection{Distorting torque}

\begin{figure}[h]
\centering{}\includegraphics[scale=0.7]{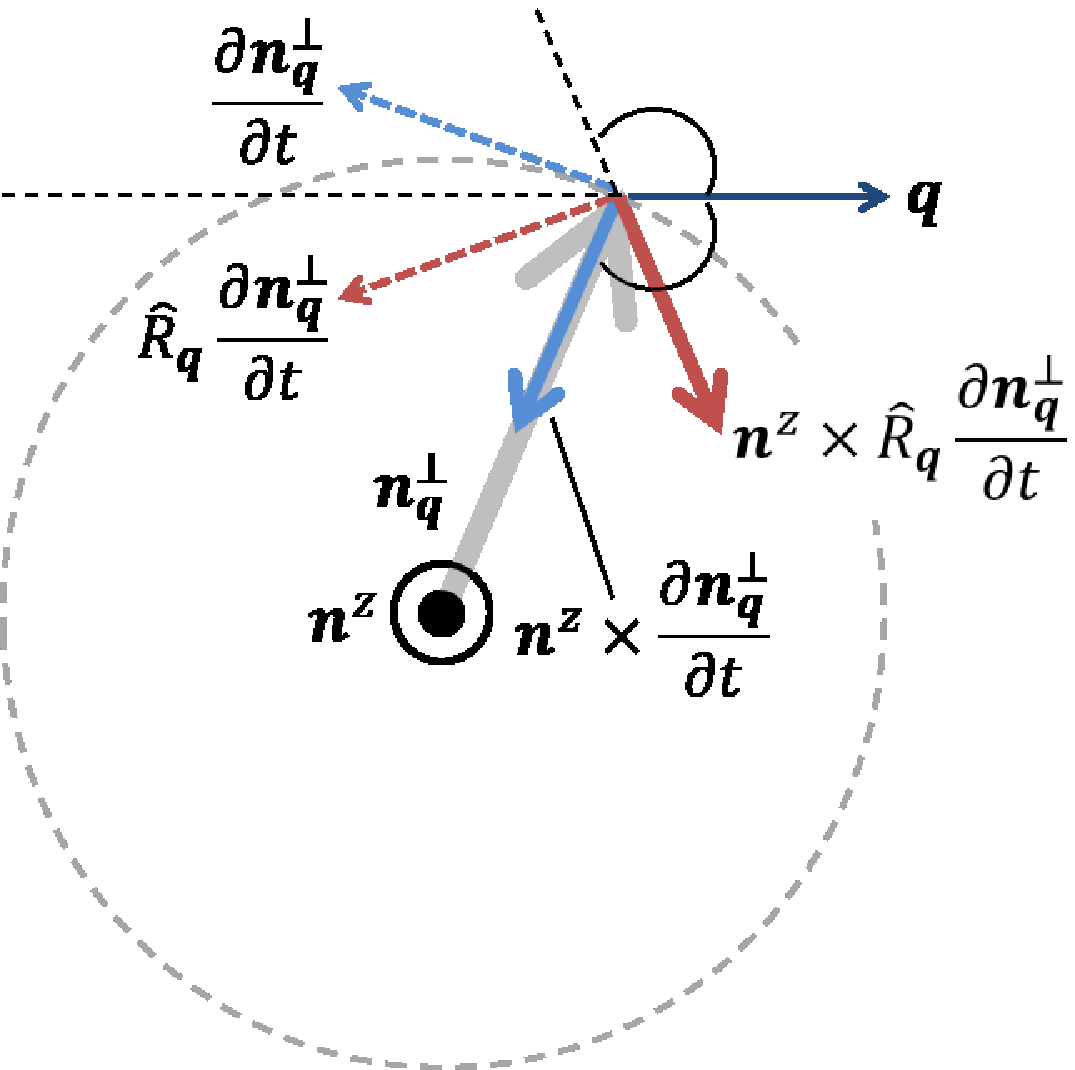}
\caption{Direction of $\bm{n}^z\times\hat{R}_{\bm{q}}\dot{\bm{n}}_{\bm{q}}^\perp$.
\label{fig:orbit1}}
\end{figure}

In this subsection, we discuss the results of both ${\alpha}^{x}_{\bm{q}}\hat{\sigma}^x$ and $\alpha^z_{\bm{q}}\hat{\sigma}^z$.
In the RSOC system, these terms vanish at $q=0$, but not at $q>0$.
From our calculations, the sum of these terms is given by
\begin{equation}
{\alpha}^{x}_{\bm{q}}\hat{\sigma}+\alpha^z_{\bm{q}}\hat{\sigma}^z
=\alpha^{xz}_{q}\hat{R}_{\bm{q}},
\end{equation}
where $\hat{R}_{\bm{q}}$ is the reflection matrix in terms of the direction $\theta_{\bm{q}}=\tan^{-1}(q_y/q_x)$:
\begin{align}
\hat{R}_{\bm{q}}=\left(\begin{array}{cc}
\cos 2\theta_{\bm{q}}
& \sin 2\theta_{\bm{q}} \\
\sin 2\theta_{\bm{q}}
& -\cos 2\theta_{\bm{q}}
\end{array}\right).
\end{align}
Moreover, $\alpha^{xz}_q$ is the $q$-dependent coefficient of new torque, $\bm{n}^z\times\hat{R}_{\bm{q}}\dot{\bm{n}}_{\bm{q}}^\perp$.
The direction of this torque is shown in Fig.\ref{fig:orbit1}.
We call this the $distorting$ torque because it applies $\bm{q}$-dependent force to the precession motion.
For example, the distorting torque is parallel to the damping torque when $\bm{n}^\perp_{\bm{q}}$ is perpendicular to $\bm{q}$;
on the other hand, the distorting torque is anti-parallel to the damping torque when $\bm{n}^\perp_{\bm{q}}$ is parallel to $\bm{q}$.

\begin{figure}[h]
\centering{}\includegraphics[scale=1.2]{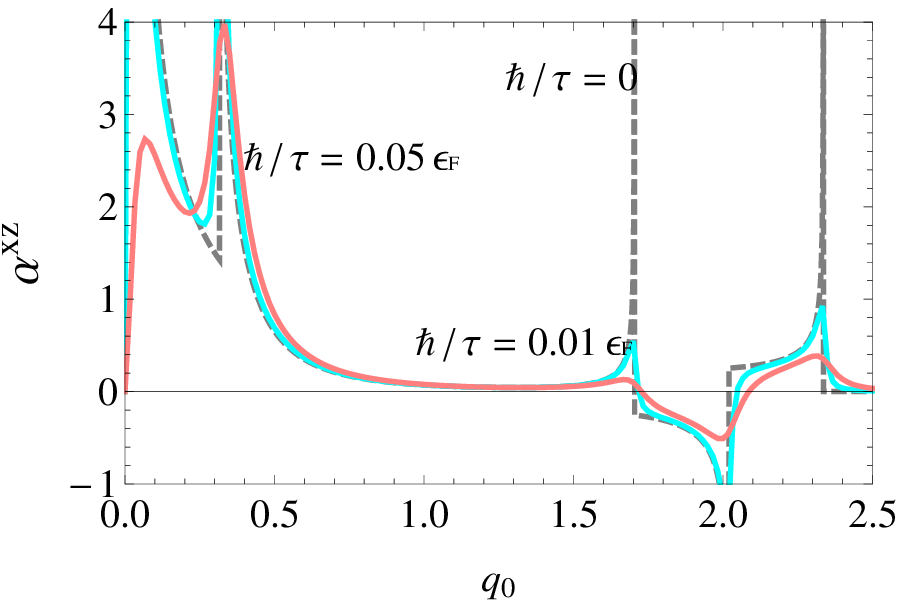}
\caption{The $q$-dependence of $\alpha_{q}^{xz}$ (divided by $\nu_{\rm F}\Delta^2/S\epsilon_{\rm F}=10^{-2}$).
The results in the clean limit are indicated by the dashed line.
\label{fig:axz}}
\end{figure}

The $q$-dependence of $\alpha_{q}^{xz}$ is shown in Fig.\ref{fig:axz}.
We confirm that the inequality equation, $\alpha_q^{xz}\leq\alpha_q$, is satisfied in the entire $q$-range.
Therefore, the precession orbits necessarily decrease with time.

In the clean limit,
\begin{equation}
\alpha^{xz}_q
=\frac{\nu_{{\rm F}}\Delta^{2}}{S\epsilon_{{\rm F}}}\lambda^{2}\sum_{\alpha\beta}\frac{\left(k_{\alpha}^{2}-k_{\beta}^{2}\right)^{2}/q_{0}^{2}+\alpha\beta\left(k_{\alpha}^{2}+k_{\beta}^{2}\right)}{\left(k_{-}^{2}-k_{+}^{2}\right)^{2}}I_{\alpha\beta}\left(q_{0}\right).
\label{eq:a1}
\end{equation}
The peaks of $\alpha_q^{xz}$ occur at same points with those of $\alpha_q$
because the right hand side of Eq.\eqref{eq:a1} includes $I_{\alpha\beta}\left(q_{0}\right)$.
This equation implies that the distorting torque is unique to the RSOC system
because $\alpha^{xz}_q$ is proportional to the square of $\lambda$.
We find that the contribution from the intra-band transitions of $\alpha_q^{xz}$,
which are given by $2\lambda^2\nu_{\rm F}\Delta^2k_\alpha^2I_{\alpha\alpha}(q_0)/S\epsilon_{\rm F}(k_-^2-k_+^2)^2$,
is equal to that of $\alpha_q$,
but the contribution from the inter-band transitions is not equal.
These results reflect the differences in the radius of the Fermi surfaces
and in the spin directions at the nesting points
between the $+$-band and the $-$-band.
These differences are attributed to the $\bm{k}$-dependent spin states originating from RSOC.
Moreover, the negative values of $\alpha_q$ at $1.7\lesssim q_0\lesssim 2$ imply
that the larger contributions come from inter-band transitions than from intra-band transitions.

\begin{figure}[h]
\centering{}\includegraphics[scale=1.2]{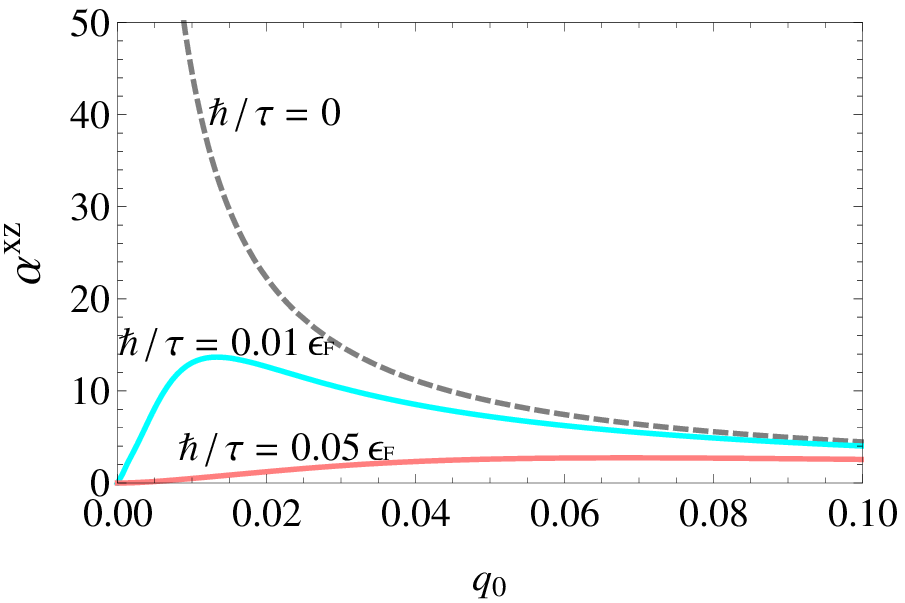}
\caption{Behavior of $\alpha_{q}^{xz}$ near $q=0$.
\label{fig:axz0}}
\end{figure}

In the presence of impurities,
the divergence of $\alpha_q^{xz}$ is suppressed similarly to the results of $\alpha_q$.
The behavior of $\alpha^{xz}_{q}$ near $q=0$ is shown in Fig.\ref{fig:axz0}.
At $q=0$, $\alpha^{xz}_q$ is always zero regardless of the presence or absence of impurities.
However, in $q\to 0$,
$\alpha_q^{xz}$ diverges (i.e., discontinuity at $q=0$) in the absence of impurities,
but $\alpha_q^{xz}$ converges to zero (i.e., continuity at $q=0$) in the presence of impurities.
In the absence of impurities, $\alpha_q^{xz}=\alpha_q$ at $q<k_--k_+$
because only intra-band transitions occur,
but this equality is no longer satisfied in the absence of impurities due to spin diffusion originating in the inter-band transitions.
Consequently, $\alpha_q^{xz}$ decreases at the peak near $q=0$, and
this peak shifts right with decreasing electron lifetimes.

\begin{figure}[h]
\centering{}\includegraphics[scale=1.2]{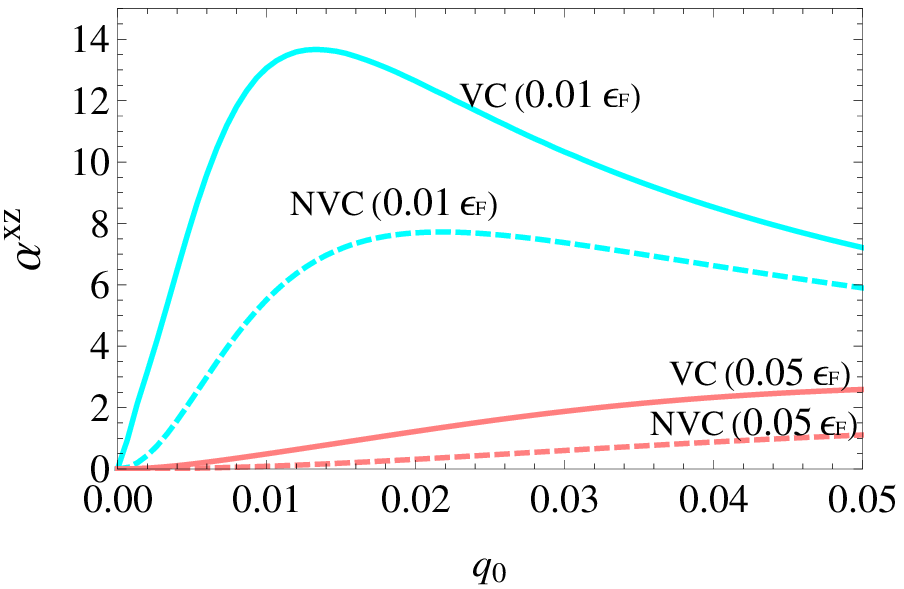}
\caption{Comparison of $\alpha_q^{xz}$ in $q_0<0.05$ including the vertex corrections (VC) and not including the vertex corrections (NVC).
The solid lines show the results for VC, and the dashed lines show the results for NVC.
The blue lines are results for $\hbar/\tau=0.01\epsilon_{\rm F}$, and
the red lines are results for $\hbar/\tau=0.05\epsilon_{\rm F}$.
\label{fig:axz_vc}}
\end{figure}

Fig.\ref{fig:axz_vc} shows the differences between the results including VC and those without VC.
The values with VC are larger than those without VC, and
VC for $\alpha_q^{xz}$ clearly has effects similar to those for $\alpha_q$ (see Fig.\ref{fig:a_vc}).

\subsection{Oscillation modes}

In the small plane wave field $\bm{h}^\perp_{\bm{q}}$,
if we can neglect $\tilde{\beta}_{\bm{q}}$, then
the linearized equation of motion is given by
\begin{equation}
\dot{\bm{n}}_{\bm{q}}^\perp=-\gamma\bm{n}_{\bm{q}}^\perp\times \bm{H}^z-\gamma\bm{n}^z\times\bm{h}_{\bm{q}}^\perp
+\bm{n}^z\times\tilde{\alpha}_{\bm{q}}\dot{\bm{n}}_{\bm{q}}^\perp.
\label{eq:linearized}
\end{equation}
We solve Eq.\eqref{eq:linearized} by assuming harmonic-time-dependence of
$\bm{h}^\perp_{\bm{q}}(\sim{\rm e}^{{\rm i}\omega t})$
without the correction term $\alpha^y_{\bm{q}}\dot{\bm{n}}^\perp_{\bm{q}}$.
We obtain the eigenequation $n_{\bm{q}}^\pm=\chi^{\pm}_{\bm{q}}h_{\bm{q}}^\pm$,
where $n^\pm_{\bm{q}}$ is the oscillation mode corresponding to each component of the magnetic field,
$h^\pm_{\bm{q}}$ ($h^+_{\bm{q}}$ is the component of right-hand rotation relative to the direction $\bm{n}^z$
and $h^-_{\bm{q}}$ is the component of the opposing direction).
Assuming $\alpha_q,\alpha_q^{xz}\ll 1$, we find
\begin{equation}
n^{\pm}_{\bm{q}}=\cos\left(\omega t\right)\pm{\rm i}\sin\left(\omega t\right)\mp{\rm i}\alpha_q^{xz}\sin\left(\omega t +\sin 2 \theta_{\bm{q}}\right),
\label{eq:elliptical}
\end{equation}
and
\begin{equation}
\chi_{\bm{q}}^\pm=\frac{\gamma}{\gamma H^z\mp {\rm i}\alpha_{q}\omega}.
\label{eq:sus+-}
\end{equation}
Eq. \eqref{eq:elliptical} indicates that each magnetization precession describes an elliptical trajectory.
Stronger SOC, which generates a large magnitude of $\alpha_q^{xz}$, produces a more elongated elliptical orbit.
When $\alpha^{xz}_{\bm{q}}>0$, the major axis of the elliptical orbit is parallel to $\bm{q}$,
while the major axis is perpendicular to $\bm{q}$ when $\alpha^{xz}_{\bm{q}}<0$.

From Eq.\eqref{eq:sus+-}, it is confirmed that $\chi^{\pm}_{\bm{q}}$ is not affected by $\alpha^{xz}_{\bm{q}}$
when $\alpha_q,\alpha_q^{xz}\ll 1$.
Therefore, $\alpha^{xz}_q$ cannot be determined experimentally in the same way as the estimation of $\alpha_q$, which can be measured from the line width of magnetic susceptibility.
To experimentally estimate the value of $\alpha^{xz}_q$,
direct observation of the elliptical orbit motion is required.

We suggest that the time-resolved magneto-optical Ker effect (TRMOKE) \cite{WalowskiKaufmannLenkHamannMcCordMuenzenberg2008}
is an appropriate method for detecting the elliptical trajectory in RSOC system.
For these observations, a large value of $\alpha^{xz}_q$ is required:
both strong RSOC and nesting effects are required.
In our results for $\hbar/\tau=0.01\epsilon_{\rm F}$,
the maximum value of $\alpha^{xz}_q$ is about $0.13$, which is sufficient for detection
at $q\simeq 0.12 k_{\rm F}\simeq O(10^{-1}){\rm \AA}^{-1}$.
However, an artificial excitation of spin wave which wave number is larger than 0.01 ${\rm \AA}^{-1}$
is technically difficult at this time.
Moreover, for clear detection of the $\bm{q}$-dependent precession motion,
the spot size of TRMOKE must be reduced from 10 $\mu$m to a few nanometers, which is currently a difficult target.
Thus, new experimental techniques are necessary to resolve these issues.

\section{Summary}

We calculate the dynamic part of the spin torque induced by the plane wave dynamics of magnetization in a RSOC system.
In addition to the conventional damping torque, our results show that
distorting torque, which is a phenomenon unique to RSOC systems,
originates from the inhomogeneous dynamics of magnetization.
The magnitudes of these torques depend on the strength of the nesting effects,
and they are reduced by decreasing the electron lifetime.
The vertex corrections for these terms are sufficiently large, especially in the long wavelength limit,
correcting  the deficiency of contributions from nesting effects, which originate in the intra-band transitions.
In the resonant plane wave field,
the oscillation mode of magnetization precession exhibits an elliptical trajectory
whose major axis depends on the direction of the wave vector.
However, to observe this elliptical precession motion,
improvements in measurement sensitivity and experimental techniques are required.


%
%

%

\begin{acknowledgments}
This work was supported by a Grant-in-Aid for Scientific Research from the Japan Society for the Promotion of Science (24-5058)  and by JSPS KAKENHI, Grant Number 25420686.
\end{acknowledgments}


%

\end{document}